\documentclass{sigchi}

\usepackage{balance}  % to better equalize the last page
\usepackage{graphics} % for EPS, load graphicx instead
\usepackage{times}    % comment if you want LaTeX's default font
\usepackage{url}      % llt: nicely formatted URLs

\makeatletter
\def\url@leostyle{%
  \@ifundefined{selectfont}{\def\UrlFont{\sf}}{\def\UrlFont{\small\bf\ttfamily}}}
\makeatother
\def\pprw{8.5in}
\def\pprh{11in}

\usepackage[pdftex]{hyperref}
\hypersetup{
pdftitle={Real-time Automatic Emotion Recognition
from Body Gestures},
pdfauthor={DIBRIS},
pdfkeywords={Body gesture analysis, motion features, motion capture systems, RGB-D cameras, machine learning, serious games},
bookmarksnumbered,
pdfstartview={FitH},
colorlinks,
citecolor=black,   
filecolor=black,
linkcolor=black,
urlcolor=black,
breaklinks=true,
}

\begin{document}

\title{Real-time Automatic Emotion Recognition
from Body Gestures}
%

%\author{Anonymous}

\numberofauthors{5}
\author{
  % 1st. author
\alignauthor Stefano Piana\\
      \affaddr{DIBRIS}\\
      \affaddr{University of Genova}\\
      \affaddr{16145 Genova, Italy}\\
      \email{stefano.piana@dist.unige.it}
\and
%% 2nd. author
\alignauthor Alessandra Staglian\`{o}\\
      \affaddr{DIBRIS}\\
      \affaddr{University of Genova}\\
      \affaddr{16145 Genova, Italy}\\
      \email{alessandra.stagliano@unige.it}
\and
%% 3rd. author
\alignauthor Francesca Odone\\
      \affaddr{DIBRIS}\\
      \affaddr{University of Genova}\\
      \affaddr{16145 Genova, Italy}\\
      \email{francesca.odone@unige.it}
\and
%% 4th. author			
\alignauthor Alessandro Verri\\
      \affaddr{DIBRIS}\\
      \affaddr{University of Genova}\\
      \affaddr{16145 Genova, Italy}\\
      \email{verri@unige.it}
%% 5th. author			
\alignauthor Antonio Camurri\\
      \affaddr{DIBRIS}\\
      \affaddr{University of Genova}\\
      \affaddr{16145 Genova, Italy}\\
      \email{antonio.camurri@unige.it}
}

% Teaser figure can go here
%\teaser{
%  \centering
%  \includegraphics{Figure1}
%  \caption{Teaser Image}
%  \label{fig:teaser}
%}

\maketitle

\begin{abstract}
Although psychological research indicates that bodily expressions
convey important affective information, to date research in emotion 
recognition focused mainly on facial expression or voice analysis. 
In this paper we propose an approach to real-time automatic emotion 
recognition from body movements.
A set of postural, kinematic, and geometrical features are extracted 
from sequences 3D skeletons and fed to a multi-class SVM classifier. 
The proposed method has been assessed on data acquired through two different systems:  a professional-grade optical motion capture system, and Microsoft Kinect. The system has been assessed on a "six emotions" recognition problem, and using a leave-one-subject-out cross validation strategy, reached an overall recognition rate of 61.3\% which is very close to the recognition rate of 61.9\% obtained by human observers. 
To provide further testing of the system, two games were developed, where one or two users have to interact to understand and express emotions with their body.

\end{abstract}

\keywords{
	Body gesture analysis, motion features, motion capture systems, RGB-D cameras, machine learning, serious games
}

\section{Introduction}
%\symbolfootnote[1]{This work has been developed inside the InfoMus Lab in Casa Paganini, Piazza di Santa Maria in Passione 34, Genoa, Italy}

In this paper we propose a method for recognizing emotional states from body motion and gestures, starting from a set of psychology inspired features extracted by 3D motion clips. We focus in particular on six archetypical emotions
(anger, happiness, sadness, disgust, fear and surprise). 

Over the years research in emotion recognition mainly focused on facial expression or voice analysis, in accordance with the intuition proposed by Ekman \cite{ekman1965differential}, who pointed out how people focus more on facial expression than body gesture when they try to understand other people's emotions.
However, recent research in experimental psychology suggests that body language constitute a significant source of affective information. Bull \cite{bull1987posture} found that interest/boredom and agreement/disagreement can be associated with different body postures/movements.
Pollick et al.\cite{pollick2001perceiving} found that given point-light arm movements human observers could distinguish basic emotions with  an accuracy significantly above the chance level.
Coulson \cite{coulson2004attributing} highlighted the role of static body postures in the recognition task where artificially generated emotional-related postures where shown to people. Techniques for the automated emotion recognition from full body movement were proposed by Camurri et al in \cite{camurri2003recognizing}.

Even more so, research in experimental psychology demonstrated how some qualities of movement are related to specific emotions: for example, the fear brings to contract the body as an attempt to be as small as possible, surprise brings to turn towards the object capturing our attention, joy may bring to openness and upward acceleration of the forearms \cite{boone1998children}. Body turning away is typical of fear and sadness; body turning towards is typical of happiness, anger, surprise; we tend to open our arms when we are happy, angry or surprised; we can either move fast (fear, happiness, anger, surprise) or slow (sadness).
In \cite{meijer1989} de Meijer presents a detailed study of how the body movements are related to the emotions. He observes the following dimensions
and qualities: Trunk movement: \textit{stretching - bowing}; Arm movement: \textit{opening - closing}; Vertical direction: \textit{upward - downward}; Sagittal direction: \textit{forward - backward}; Force: \textit{strong - light}; Velocity: \textit{fast - slow}; Directness: \textit{direct - indirect}.
de Meijer notices how, various combinations of those dimensions and qualities can be found in different emotions. For instance a joyful feeling could be characterized by a Strong force, a fast Velocity and a direct Trajectory but it could have a Light force as well, or be an indirect movement. 

Works inspired by these studies that involve analysis of the entire body's movement showed that movement-related dimensions can be used in the inference of distinct emotions \cite{meijer1989,wallbott98,camurri2003,glo2011}.

Kapoor et al. \cite{kapoor2007automatic} demonstrated a correlation between body posture and frustration in a computer-based tutoring environment.
In a different study Kapur et al. \cite{kapur2005gesture} showed how four basic emotions could be automatically distinguished from simple statistical measures of motion's dynamics.
Balomenos et al. \cite{balomenos2005emotion} combined facial expressions and hand gestures for the recognition of six prototypical emotions.

In this work we focus on  {\em automatic emotions recognition} starting from an automatic analysis of body movements; our research is based on the studies from experimental psychology mentioned previously (de Meijer\cite{meijer1989}, Wallbott \cite{wallbott98}, Boone \& Cunningham \cite{boone1998children}) and from humanistic theories (Laban effort \cite{Laban47,Laban63}).

In our study we start from detailed 3D motion data recordings of full-body movements obtained by professional grade optical motion capture systems (e.g., Qualysis\cite{Qualisys}) and video cameras. 

Also, we assess the appropriateness of our method on 3D data acquired by  low-cost RGB-D sensors (e.g., Kinect\cite{Kinect}).
The adoption of low-cost measuring devices, which are less precise but easier to acquire and to install, will enable us to integrate our methods with serious games supporting autistic children to learn to recognize and to express emotions, which is one of the main goals of the EU \textit{ASc-Inclusion}  Project which is funding the study.

We then extract features of body movement, most of them derived by psychology studies. These features are integrated over time and then used them to build a feature vector of a movement (or gesture) portion which is fed to a machine learning classifies.

The paper is organized as follows: the Framework Section introduces emotion-related dimensions, describes the data representations and the classification procedure; the Experiments Section describes the experimental set-up, the validation process, and two applications of the framework; finally, conclusions are given and future work on this topic is described.

\section{The Framework}
\label{sec::Framework}
The work described in this paper is part of the \textit{Anonymous} Project, that aims to develop ICT solutions to assist children affected by Autism Spectrum Conditions (ASC). It focuses on the development of serious games to support children to understand and express emotions. The complete framework will process facial expressions (see \cite{golan2006cambridge}), voice (see \cite{golan2010enhancing}), and full-body movements and gestures. The stand-alone system described in this paper will be integrated with the other two modalities for the final version of the serious games. 

The set of emotions considered in this work is limited to six emotions (Happiness, Anger, Sadness, Surprise, Fear, and Disgust), because they are widely accepted and recognized as "basic" (especially from the body gestures point of view) and cross-cultural (see Cornelius \cite{cornelius1996science} and \cite{cornelius2000theoretical}).

\subsection{The computed dimensions}
It is well-known how the body is able to convey and communicate emotions and in general implicit information. 
Starting from the tracking of joints positioned on the upper body (head, shoulders, elbows, hands, and torso) a set of relevant low- and mid-level movement related features was identified and different algorithms were implemented to extract these features (see \cite{piana2013set}).
The joints were chosen according to the study of Glowinsky et al. (\cite{glowinski2011toward}), and extracted with two different modalities, as described in the experiments Section.
In particular we focused on the set of dimensions described below:
\begin{itemize}
\item {\em From all the tracked  joints:}
\begin{itemize}
	\item Kinetic Energy, Contraction Index/Contraction-Expansion, Periodicity, Global direction of motion, Overall symmetry
\end{itemize}
\item {\em From the head joint}:
\begin{itemize}
	\item Speed, Acceleration, Jerk, Leaning speed, Leaning position, Symmetry w.r.t. hands, Symmetry w.r.t. shoulders
\end{itemize}
\item {\em From the torso joint}:
\begin{itemize}
	\item Leaning speed, Leaning position
\end{itemize}
\item {\em From the hand(s) joint(s)}:
\begin{itemize}
	\item Distance between hands, Smoothness, Speed, Acceleration
	 Jerk,
	 Periodicity,
	 Direction,
	 Fluidity,
	 Impulsiveness,
	 Curvature of the trajectory,
	 Distance from the torso joint,
	 Distance from the head.
\end{itemize}
\item {\em From the shoulders or elbows joints}:
\begin{itemize}
	\item Speed,
	 Acceleration,
	 Jerk,
	 Fluidity,
	 Direction,
	 Periodicity,
	 Curvature of the trajectory,
	 Distance from the torso (elbows only)
\end{itemize}
\end{itemize}  

\subsection{Data representation and classification}

The body joints are tracked and the resulting movements are segmented into single gestures, according to threshold over some features, e.g. low Kinetic Energy. For each gesture, we build a time series per each feature. In general time series will have variable lengths, depending on the different speed of the gesture and its complexity. At the same time, from the machine learning stand point, we assume that points are feature vectors leaving in a fixed $n-$dimensional space. Therefore it is common practice to map initial representations on an intermediate (fixed length) representation \cite{noceti2011learning, noceti2012learning}. There are different ways of doing so, ranging from a sub-sampling of variable length series with a fixed number of samples (but in this case we may miss important information), to clipping a time series into a subset of fixed length sub-sequences (and in this case the size of the sub-sequences will be a crucial parameter). Alternatively one may resort to alternative descriptions, such as first or second order statistics. In this work, after a preliminary experimental analysis, we rely on first order statistics and compute histograms of each feature observed over time. Once an appropriate quantization is chosen, all histograms will have a fixed length. In this way we lose the data temporal coherence (which will be investigated in future), in favour of a more compact representation. 
\begin{figure*}
\centering
\begin{tabular}{cc}
\includegraphics[width=0.4\textwidth]{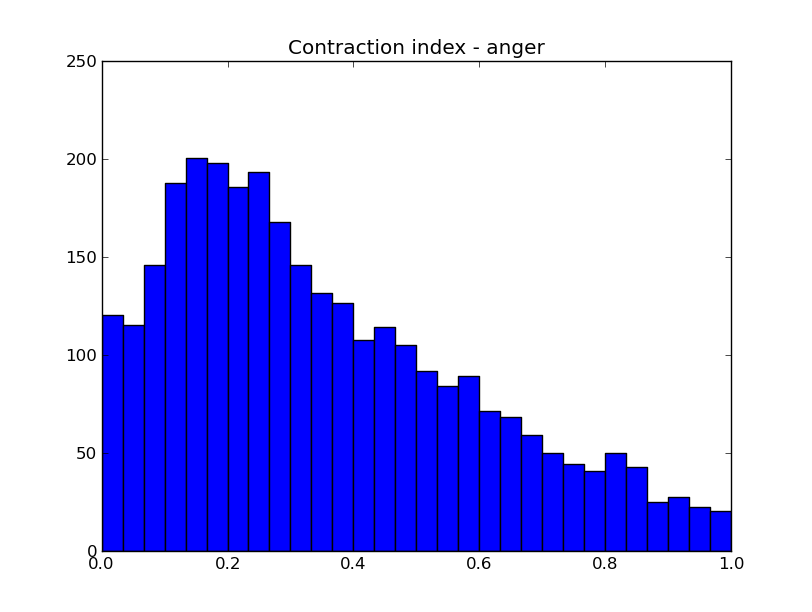} &
\includegraphics[width=0.4\textwidth]{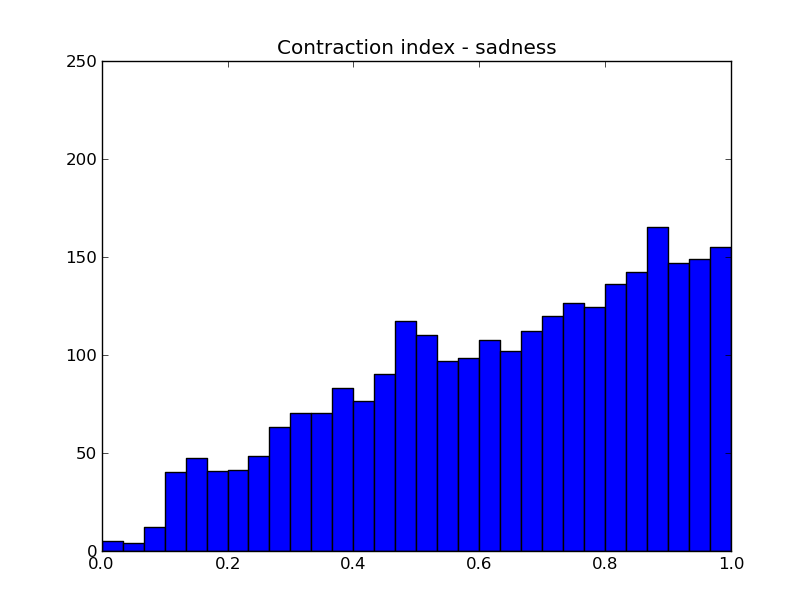} \\

\includegraphics[width=0.4\textwidth]{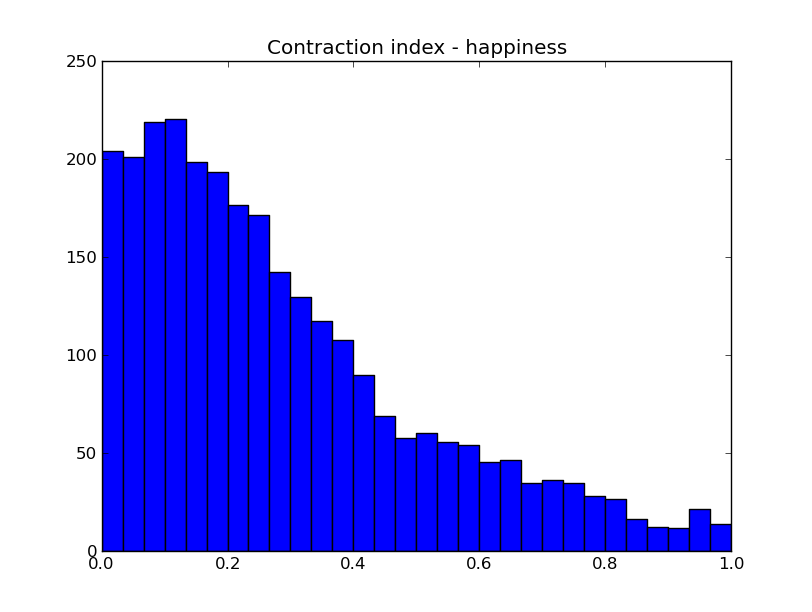} &
\includegraphics[width=0.4\textwidth]{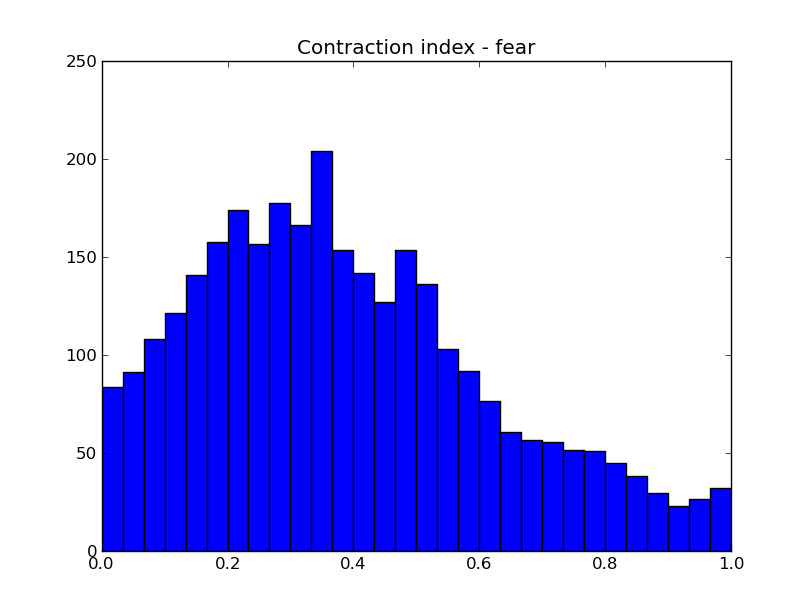} \\

\includegraphics[width=0.4\textwidth]{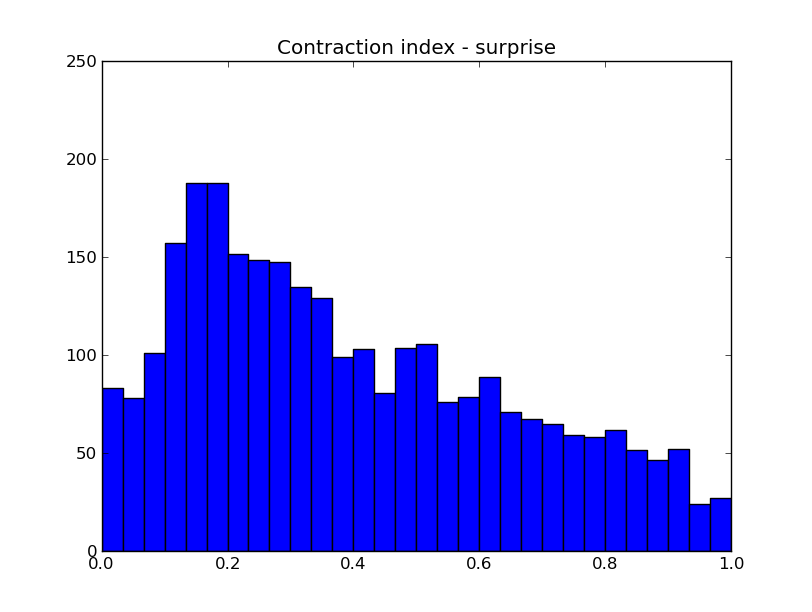} &
\includegraphics[width=0.4\textwidth]{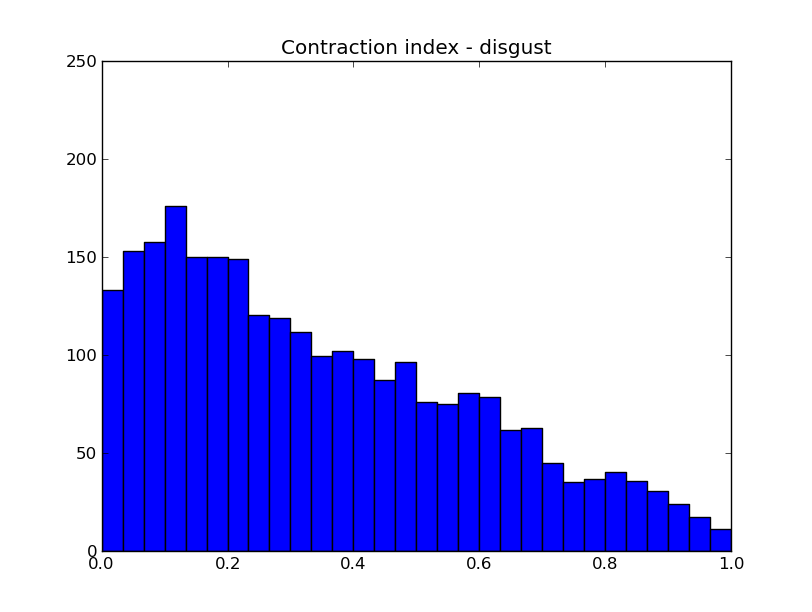} \\
\end{tabular}

\caption{\label{FigHistExample} 30 bins cumulative histograms of the Contraction Index of 100 segments per emotion produced from the recordings of 12 people: anger (top left), sadness (top right), happiness (middle left), fear (middle right), surprise (bottom left), and disgust (bottom right).}
\end{figure*}

Figure \ref{FigHistExample} shows  cumulative histograms of the kinetic energy of 100 segments for each one of the six basic emotions. The histograms have different characteristics, meaning that the kinetic energy takes different values in the different emotions, and this information can be used to discriminate between them.
In practice, a given gesture is represented by the concatenation of the histograms computed for each feature forming a high dimensional feature vector. Such feature vectors are used for the classification step.

Machine Learning is used to discriminate between the different emotions. After the representation step, a one-versus-one linear SVM \cite{cortes1995support} classifier has been trained and an ECOC (Error Correcting Output Coding - \cite{klautau2002combined}) procedure was used to refine the results of the overall multi-class classifier.

\section{Method assessment}
\label{sec::Experiments}

We collected a dataset of people expressing emotions with their body, using two different acquisition modalities: the Qualisys motion capture system \cite{Qualisys}, and a Microsoft Kinect \cite{Kinect}.  The two acquisition systems are very different and are indeed meant for two different purposes: Qualisys has been used at an early stage of development, as a proof of concept of the proposed methodology, Kinect has been adopted later to test the applicability of the method in real-world applications and games.

Qualisys is a sophisticated motion capture system, composed by different (9 in our setup) infrared cameras that capture the light reflected by markers. The data it provides are precise and reliable, but as a set up it has many drawbacks: as a first thing it is very expensive, it requires a complex calibration procedure and a large space (and, for these reasons, it has a limited portability). Finally, it requires the user to wear markers, making it inappropriate for users with Autism Spectrum conditions.
Instead, Microsoft Kinect as a commercial product is very easy to acquire and to install, it has very few requirements on the acquisition environment, but provides noisier data.

In both cases we recorded sequences of 3D coordinates (or 3D skeletons), corresponding to the same body joints.

The two datasets include 12 actors, of whom four were female and eight males, aged between 24 and 60, performing the six basic emotions with their body, each of them for a number of times ranging from  three to seven. We obtained a total of about 100 videos, which have been segmented manually to separate clips (or segments) of expressive gesture. Each datum has been associated with a label, obtained by the actor stating what type of emotion he or she was expressing. These labels form the datasets ground truth. 

\subsection{Humans Data Validation}
To evaluate the degree of difficulty of the task, the obtained segments have been validated by humans (60 anonymous people who did not participate to the data acquisition process) through an on-line survey. 

\begin{figure}
\centering
\includegraphics[width=0.45\textwidth]{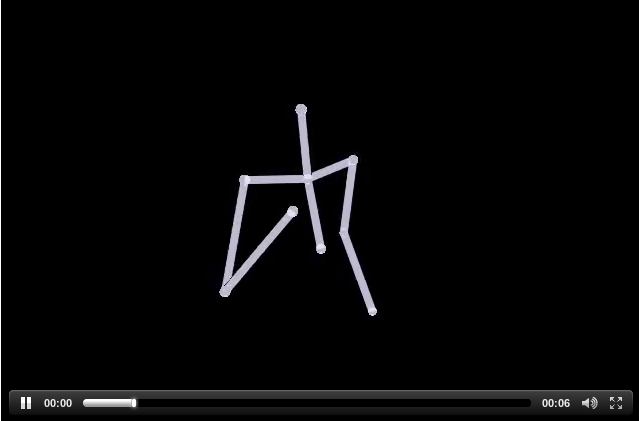}
\caption{\label{skeleton} An example of the stimuli used for data validation with humans.}
\end{figure}

Each person was shown 10 segments of the 3D skeletons, and they had to guess which emotion was being expressed or, alternatively, choose a "I don't know" reply. An example of the input stimulus provided is shown in Figure \ref{skeleton}, where the limited amount of information conveyed by the data is apparent. Indeed, the goal of this experiment was to understand to what extent a human observer is able to understand emotions simply by looking at body movements. The sole 3D skeleton is a guarantee that the user is not exploiting other information, such as insights from facial expressions or contextual information. 
The human validation underlined the fact that three out of the six basic emotions (happiness, sadness, and anger) were clearly recognizable from body movements, while the other three (surprise, disgust, and fear) were easily confused with one another. For this reason a sub-problem of four classes (happiness, sadness, anger, and fear) has also been taken into account.
The results of the human validation, with respect to the data ground truth, are reported in Table \ref{TableValidation} and clearly testify the difficulty of the problem we are addressing. 

\begin{table}
\centering
\begin{tabular}{ccc}
\hline
& 6 emotions & 4 emotions \\
\hline
happiness & 81,3\% & 87,5\% \\
fear & 48,5\% & 81,2\% \\
disgust & 37,2\% &  -  \\
sadness & 86,7\% & 94,9\% \\
anger & 73,9\% & 82,0\% \\
surprise & 35,2\% &  -  \\
\hline
average & 61,9\% & 85,2\% \\
\hline

\end{tabular}
\caption{\label{TableValidation} Human validation results. First column: name of the emotion. Second column: results obtained with 6 different emotions. Third column: results obtained with 4 different emotions.}
\end{table}

\subsection{Classification results}
Two different problems have been considered: the first one considering only four emotions (Happiness, Sadness, Anger, Fear), and the second considering all of the six basic emotions. 

The two problems were assessed with two different type of data: a smaller, but with more precise measurements, dataset recorded with the Qualisys MoCap system, and a bigger dataset containing data recorded with the Kinect sensor. For the four classes problem, the Qualisys data were composed of 213 and the Kinect of 398 gestures, while for the six classes problems there were 310 and 579 gestures. This difference in the number of data is due to the fact that it was much more difficult to record people with the Qualisys system. 

The emotion dimensions described in Section "The Framework" have been computed from the data and summarized in 30-bins  histograms. Then a linear one-vs-one SVM with ECOCs has been used to classify the data. 

A first set of experiments has been performed by splitting the dataset randomly in a training and a test set (the procedure was repeated 50 times, the results we show are an average of the repetitions): in this case, gestures performed by the same subject might appear both in the training and in the test set.
A second set of experiments has been carried out with a Leave-One-Subject-Out cross validation (LOSO cv), training the classifiers over 11 subjects and testing them with data of the 12th left out subject. 
\begin{table}
\centering
	\begin{tabular}{ccc}
		\hline
     & Split Data & LOSO cv \\
     \hline
	happiness & 65.2\% & 61.1\%   \\
	anger & 78.6\% & 74.5\% \\
	sadness & 90.3\% & 85.4\%  \\
	fear & 82.5\% & 77.5\% \\
	\hline
	total & \textbf{79.2\%} & \textbf{74.6\%} \\
	\hline
	\end{tabular}
\caption{\label{Table4clQ} Qualisys data: classification results of the 4 classes problem.}
\end{table}
\begin{table}
\centering
	\begin{tabular}{ccc}
		\hline
     & Split Data & LOSO cv \\
     \hline
	happiness & 62.1\% & 44.4\%   \\
	anger & 71.1\% & 68.6\% \\
	sadness & 77.2\% & 77.1\%  \\
	fear & 55.3\% & 51.7\% \\
	disgust & 53.5\% & 48.8\% \\
	surprise & 54.7\% & 34.6\% \\
	\hline
	total & \textbf{62.3\%} & \textbf{54.2\%} \\
	\hline
	\end{tabular}
\caption{\label{Table6clQ} Qualisys data: classification results of the 6 classes problem.}
\end{table}
\begin{table}
\centering
	\begin{tabular}{ccc}
		\hline
     & Split Data & LOSO cv \\
     \hline
	happiness & 74.7\% & 58.1\%   \\
	anger & 86.2\% & 76.6\% \\
	sadness & 93.4\% & 88.3\%  \\
	fear & 86.5\% & 87.8\% \\
	\hline
	total & \textbf{85.2\%} & \textbf{77.7\%} \\
	\hline
	\end{tabular}
\caption{\label{Table4cl} Kinect data: classification results of the 4 classes problem.}
\end{table}
\begin{table}
\centering
	\begin{tabular}{ccc}
		\hline
     & Split Data & LOSO cv \\
     \hline
	happiness & 59.8\% & 47.6\%   \\
	anger & 75.3\% & 70.2\% \\
	sadness & 85.7\% & 80.2\%  \\
	fear & 71.1\% & 60.7\% \\
	disgust & 75.8\% & 57.1\% \\
	surprise & 63.7\% & 52.2\% \\
	\hline
	total & \textbf{71.9\%} & \textbf{61.3\%} \\
	\hline
	\end{tabular}
\caption{\label{Table6cl} Kinect data: classification results of the 6 classes problem.}
\end{table}

Table \ref{Table4clQ} and Table \ref{Table6clQ}  report the classification results obtained with Qualisys, while Table \ref{Table4cl} and Table \ref{Table6cl}  report the ones obtained with the Kinect data. In general, the accuracy is higher if the data are split randomly. This was expected mainly because of the limited size of the datasets for a problem of large complexity and variability: different people may express emotions in slightly different ways, therefore having examples from each subject both on training and testing is beneficial for the results. To contrast the likelihood of over fitting we are working on acquiring new data and, more importantly, enlarge the number of available actors: this would probably lead to an improvement of the overall accuracy in the LOSO cv experiments. 
The importance of a large dataset is also confirmed by the fact that Qualisys data, albeit more accurate, lead to a lower accuracy.

Comparing the results with the human validation data we notice how the system achieves the same level of accuracy as humans.

\section{Application to the design of serious games}

%preso paro paro (o quasi) dal deliverable
The automatic emotion recognition system has been used as a building block in the design of serious games developed within the \textit{ASC-Inclusion} project. 

The body movement analyser is based on the EyesWeb XMI development platform \footnote{ http://www.infomus.org/eyesweb   }, that allows  us to extract the feature vectors in real-time, and to give a feedback to the user.

Our goal is to implement and evaluate an on-line body expression analyser, showing children with Autism Spectrum Condition how 
they can improve their context-dependent body emotional expressiveness. Based on the child's movement input recorded by RGB-D sensors (as the Microsoft Kinect), the set of emotionally relevant features is extracted and analysed to infer the child's emotional state.

The system described in this paper was used as the main part of two game demonstrations. Both the games perform a live automatic emotion recognition, and interact with the user by asking to guess an emotion and to express an emotion with the body. In the remainder of this section the GUIs and the games will be described in more details.

\subsection{Graphical User Interface}
\begin{figure}
\centering
\includegraphics[width=0.45\textwidth]{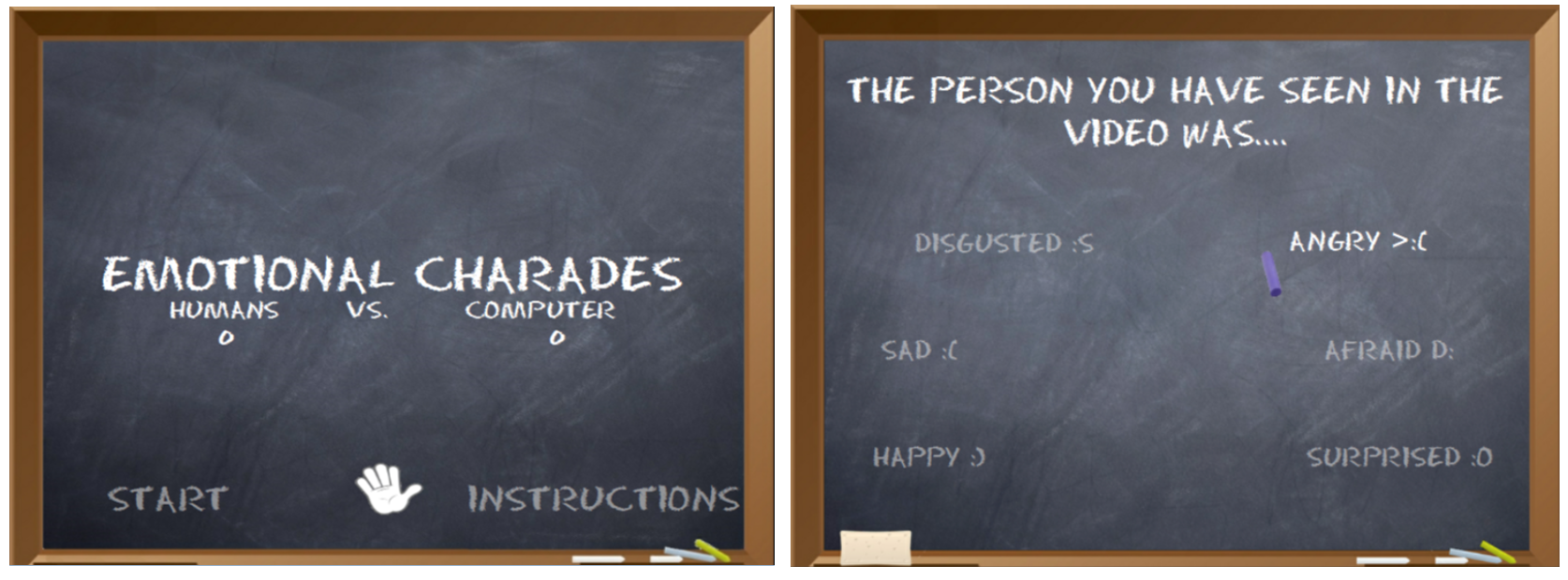}
\caption{\label{FigGuiExample} Two different screens of the GUI.}
\end{figure}

Both the game demonstrations share a very simple GUI that recalls a blackboard with white chalk writings over it. The interaction is performed with a Kinect: the user has to select different buttons driving a pointer with a hand. There is no sound, all the questions and the instructions are made through text. Figure \ref{FigGuiExample} shows examples of the GUI. 
\subsection{Body Emotion Game Demonstration}
This is a short game that is composed of two parts.
In the first part the system shows to the player a video of a person that expresses an emotion. The video stresses the information carried by the body movements and discard other stimuli (context, facial expression, voice). 
Then, the system shows a set of possible emotions (the six basic emotions: sadness, anger, happiness, fear, disgust, and surprise) and the player has to select the emotion that in her opinion the person in the video was feeling. If the player selects the correct emotion she gains one point. 
In the second part the player has to act the same emotion of the video with the body. The system will try to understand what emotion is being expressed by the player. If the recognized emotion is the correct one, the player gains another point. If the recognized emotion is different from the correct one, the system will ask to the player if she wants to try again. If she says no, she will not gain any other point. 
She will have up to three attempts, if the system answers correctly, or the user spends all the three attempts, the game ends and the user is given a final score that includes the points acquired during the two phases of the game.

\subsection{Emotional Charades}
\begin{figure}
\centering
\includegraphics[width=0.45\textwidth]{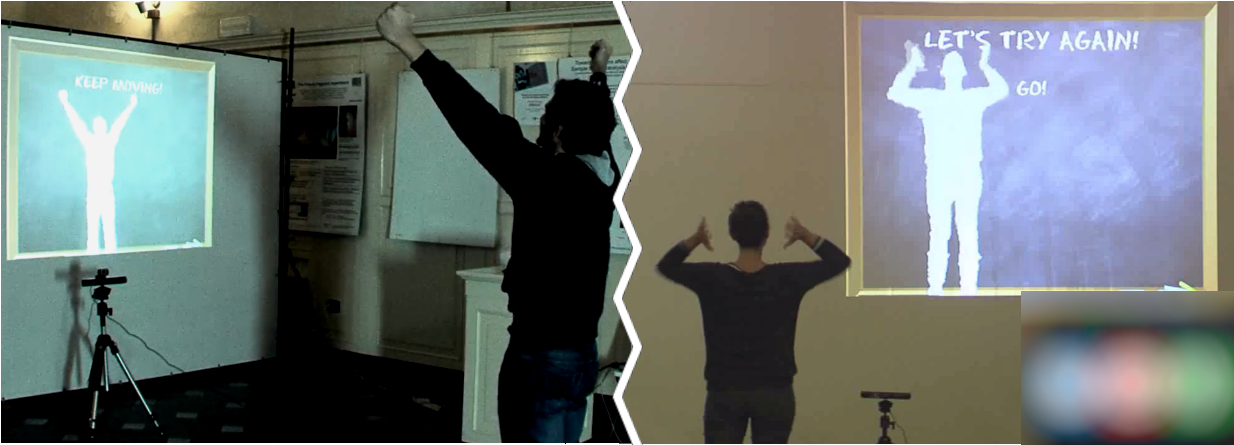}
\caption{\label{GameExample} An example of interaction between two players}
\end{figure}

This is a two-players game. 
Player 1 and Player 2 cannot see each other, they both have a computer with a Kinect. In the first round, Player 1 has to choose one of the six base emotions and express it with the body. Player 2 will see only the depth map produced by Player 1 and will have to guess which emotion was chosen by her. The computer will try to do the same. 
Player 1 will be shown the answers given by the computer and Player 2, and will say if they have guessed or not. If both have guessed, Player 1 gains 2 points and Player 2 - 1 point, if only one of them is right Player 1 will gain 1 point (if only Player 2 is right she will gain 2 points), if neither the computer nor Player 2 are correct, Player 1 won't gain any point while Player 2 will gain 1 point. 
The game continues with the roles inverted. An extra score is shown at the beginning of the game, that counts how many correct answers were given by the human players and how many were given by the computer. Figure \ref{GameExample} shows an example of interaction between two players.

\section{Conclusions and Future Work}
\label{sec::Conclusions}

In this work we have presented a complete framework to monitor and process body movements to understand emotions. 
Machine Learning techniques have been applied to data extracted in real-time to interact with a human being. 
The system relies on a motion capture device, Python libraries, and EyesWeb XMI. 
The obtained classification results are very encouraging, compared to the ones obtained by the human beings validation. 

The next steps will concern both the data representations and the feedback given by the system. Further adaptive data representations, such as Dictionary Learning, will be investigated to see if the accuracy can improve. 
From the discrete classification labelling, the system will move to a continue Valence-Arousal labelling, to be more coherent with the rest of the platform of the \textit{ASC-Inclusion} project.

\bibliographystyle{acm-sigchi}
\bibliography{sigproc}
\end{document}